\documentclass[shortauth]{aa}  

\usepackage{listings} 
\usepackage{pythonhighlight} 

\usepackage{color}
\usepackage[colorlinks,breaklinks]{hyperref}
\hypersetup{linkcolor=blue,citecolor=blue,filecolor=black,urlcolor=blue}
\usepackage{amssymb}

\newcommand{\mr}[1]{{#1}}

\newcommand{\slobject}[1]{\texttt{#1}}
\newcommand{\instance}[1]{\texttt{#1}}
\newcommand{\method}[1]{\texttt{#1}}
\newcommand{\attribute}[1]{\texttt{#1}}

\usepackage{xspace}

\usepackage{graphicx}
\usepackage[utf8]{inputenc}
\usepackage{txfonts}
\usepackage{orcidlink}

\usepackage[switch]{lineno}

\begin{document}

 \title{\texttt{slicersim}: A python package to simulate \\ image slicer spectroscopic observations --- \\ application to the Lazuli Spectrograph.}
   \author{
    Rigault, M.\inst{\ref{ip2i}}\fnmsep\thanks{\texttt{m.rigault@ip2i.in2p3.fr}} \orcidlink{0000-0002-8121-2560},
    Copin, Y.\inst{\ref{ip2i}},
   Aldering, G.\inst{\ref{lbnl}},
   Edelstein, J.\inst{\ref{spaceberkeley}},
   Furesz, G.\inst{\ref{ssci}},
   Lasker, J.\inst{\ref{ssci}},
   Miller, T.\inst{\ref{spaceberkeley}} \orcidlink{0009-0005-2445-7544}, \\
   Perlmutter, S.\inst{\ref{lbnl}, \ref{ucb}},
   Roy, A.\inst{\ref{ssci}} \orcidlink{0000-0001-8127-5775},
   Schlawin, E.\inst{\ref{ssci}},
   Stefansson, G.\inst{\ref{ssci}, \ref{anton}}
    }
   \institute{
   Université Lyon 1, CNRS, IP2I Lyon, UMR 5822, Villeurbanne, France 
   \label{ip2i}
   \and
   Physics Division, Lawrence Berkeley National Laboratory, One Cyclotron Rd., Berkeley, CA 94720, USA
   \label{lbnl}
   \and
   Space Sciences Laboratory, University of California, Berkeley, Berkeley, CA 94720, USA
   \label{spaceberkeley}
   \and
   Astrophysics \& Space Institute, Schmidt Sciences, New York, NY 10011, USA
   \label{ssci}
   \and
   Physics Department, University of California, Berkeley, CA 94720, USA
   \label{ucb}
   \and
   Anton Pannekoek Institute for Astronomy, 904 Science Park, University of Amsterdam, Amsterdam 1098 XH, The Netherlands
   \label{anton}
   }
\titlerunning{slicersim}
\authorrunning{M. Rigault et al.}

\abstract
{Integral Field Spectroscopy (IFS) is a key technique to study galaxies, stellar and planetary systems, and transients, by spatially splitting the scene and dispersing this light onto the detector. 
This technique is now central for many modern telescopes, including the Lazuli Space Observatory which will host a low spectral-resolution image slicer spectrograph.}
{To prepare for future IFS-equipped instruments and to better understand the systematics of current ones, realistic simulations are needed.}
{We present \texttt{slicersim}, a Python package designed to simulate IFS observations. \texttt{slicersim} is a modular and flexible tool that allows the user to build a scene, define a telescope, a spectrograph, and a detector, and to simulate the resulting observation, including various sources of noise and instrumental effects.}
{The package is designed to be easily extensible to any IFS, with a current focus on that from the Lazuli Space Observatory. In this paper, we present the structure and logic of the package, detailing its main components, and providing an example of its usage.} 
{Using \texttt{slicersim} we demonstrate the feasibility of a large cosmology campaign on Lazuli --- a sample of 8000 Type~Ia Supernovae uniformly distributed between $z=0$ and $z=1.5$, observed within the first few years of its operations for a typical average signal-to-noise of 25 per resolution element \mr{rest-frame optical} wavelengths.}


   \maketitle
   
\section{Introduction}

For decades, the study of astrophysical transients, such as supernovae (SNe), has been based on the combination of an initial discovery emerging from a photometric image survey, enriched by follow-up data, including more precise photometric data and spectroscopic information. The latter has been particularly important in analyzing those transients, either to (sub)classify them \citep[e.g.,][]{filippenko1997,folatelli2013, dimitriadis2025} or to study their properties and those of their progenitor systems \citep[e.g.,][]{blondin2012, silverman2012, galyam2014,maguire2014, burgaz2025}. 

In the last few decades, photometric discovery rates of SNe have surpassed the spectroscopic follow-up capabilities \citep[see discussion in][]{kulkarni2020}.
This led the Dark Energy Survey \citep[DES,][]{des2024} 
perform the first state-of-the-art Type~Ia supernova cosmology analysis using a sample of $\mathcal{O}(1,000)$ Type~Ia Supernovae (SNe\,Ia) that were classified exclusively using photometry to probe cosmological parameters \citep{moller2020, boone2021}.
However, it is now known that the right spectroscopic information can provide higher precision (lower residual scatter) and higher accuracy (e.g., lower astrophysical biases) SN\,Ia distances \citep{boonetwin2021, stein2022, ganot2025}.

The advent of wide-field imaging surveys such as the Zwicky Transient Facility \citep[ZTF,][]{bellm2019, graham2019}, the Vera C. Rubin Observatory's Legacy Survey of Space and Time \citep[LSST,][]{ivezic2019} and the upcoming Nancy Grace Roman Space Telescope \citep[\textit{Roman},][]{spergel2015} vastly increases the rate of photometric transient discovery and further challenges the spectroscopic follow-up capacities. 
Thanks to its dedicated low-resolution integral field spectrograph \citep[the $R\sim100$ SED machine,][]{blagorodnova2018, rigault2019} ZTF has demonstrated the ability to complete an $\mathcal{O}(10,000)$ spectroscopically classified magnitude limited survey \citep[][]{fremling2020,perley2020}, including more than 1000 cosmology-grade SN~Ia per year \citep{rigault2025}. This illustrates the interest in such low-resolution spectroscopy for supernova physics and cosmology. The use of low spectral resolution avoids the need for multiple wavelength channels and for faint sources reduces the exposure time needed to reach a significant signal-to-noise per fixed wavelength interval, but remains high enough to capture wide spectroscopic features associated with explosive events such as supernovae over the wide wavelength range where such features occur. Furthermore, an integral field spectrograph (IFS) design simplifies target acquisition, mitigates slit losses, enables the more accurate 3D-PSF source extraction, while enabling improved host galaxy background subtraction \citep{bongard2011,rigault2019,lezmy2022}, and thereby allows for spectrophotometric calibration of the instrument \citep{rubin2022}. 


IFSs are increasingly being developed in astronomy following success of first instruments such as TIGER  \citep{courtes1982, bacon1995}, SAURON \citep{bacon2001}, SNIFS \citep{aldering2002b}, MUSE \citep{bacon2010}, and MaNGA \citep{drory2015} among many others. Most recently, the James Webb Space Observatory has flown two (MIRI and NIRSpec; \citealt{glasse2015,jakobsen2022}) and HARMONI \citep{thatte2021} will be a first generation instrument for the upcoming Extremely Large Telescope. Furthermore, the Lazuli Space Observatory \citep{roy2026} has been designed to host a low-resolution IFS to follow-up, among other goals, LSST and \textit{Roman} deep fields transients, such as SNe\,Ia. As presented in \cite{roy2026}, the optical design of the Lazuli IFS builds on slicer integral field unit (IFU) designs from SNAP \citep{aldering2002a, ealet2006} and ORKID-II \citep{pasquale2024}.

In this context, we introduce \texttt{slicersim}, a Python package to simulate IFS observations.
The \texttt{slicersim} package has been designed to be a flexible and modular API.
It is structured to be easy to use while allowing the user to build complex observations. This package has been developed to support the Lazuli Space Observatory \citep{roy2026}, which hosts a slicer-spectrograph, but its code structure is generic enough to be adapted to any kind of IFS. As such, this package could be used to help aid the conceptual design of instruments and observation campaigns, as we show for Lazuli, and to test data processing pipelines and train simulation based inference networks.

This paper is structured as follows. 
In Section~\ref{sec:structure}, we present the structure and logic of the package. 
In Section~\ref{sec:model}, we detail the model and the assumptions currently implemented in \texttt{slicersim}. In that section, we also provide examples of the package usage, illustrating how one can use it to get a simple exposure time or to access details about instrumental, telescope, and astrophysical contributions to the variance.
We also show that all simulation parameters can be changed on the fly to study their impact on the observations.
We conclude and discuss future developments in Section~\ref{sec:conclusion}.

This paper presents \texttt{slicersim} version 1.0 (\texttt{v.1.0.0}). Future development may change part of the code and we invite the user to refer to the latest \href{https://skysurvey.readthedocs.io/en/latest}{online documentation}. The code is publicly available on \href{https://github.com/schmidt-observatories/slicersim}{GitHub} and can be installed using \texttt{pip install slicersim}.

\section{Package Structure and Logic}
\label{sec:structure}

\begin{figure*}
  \sidecaption
  \includegraphics[width=0.7\linewidth]{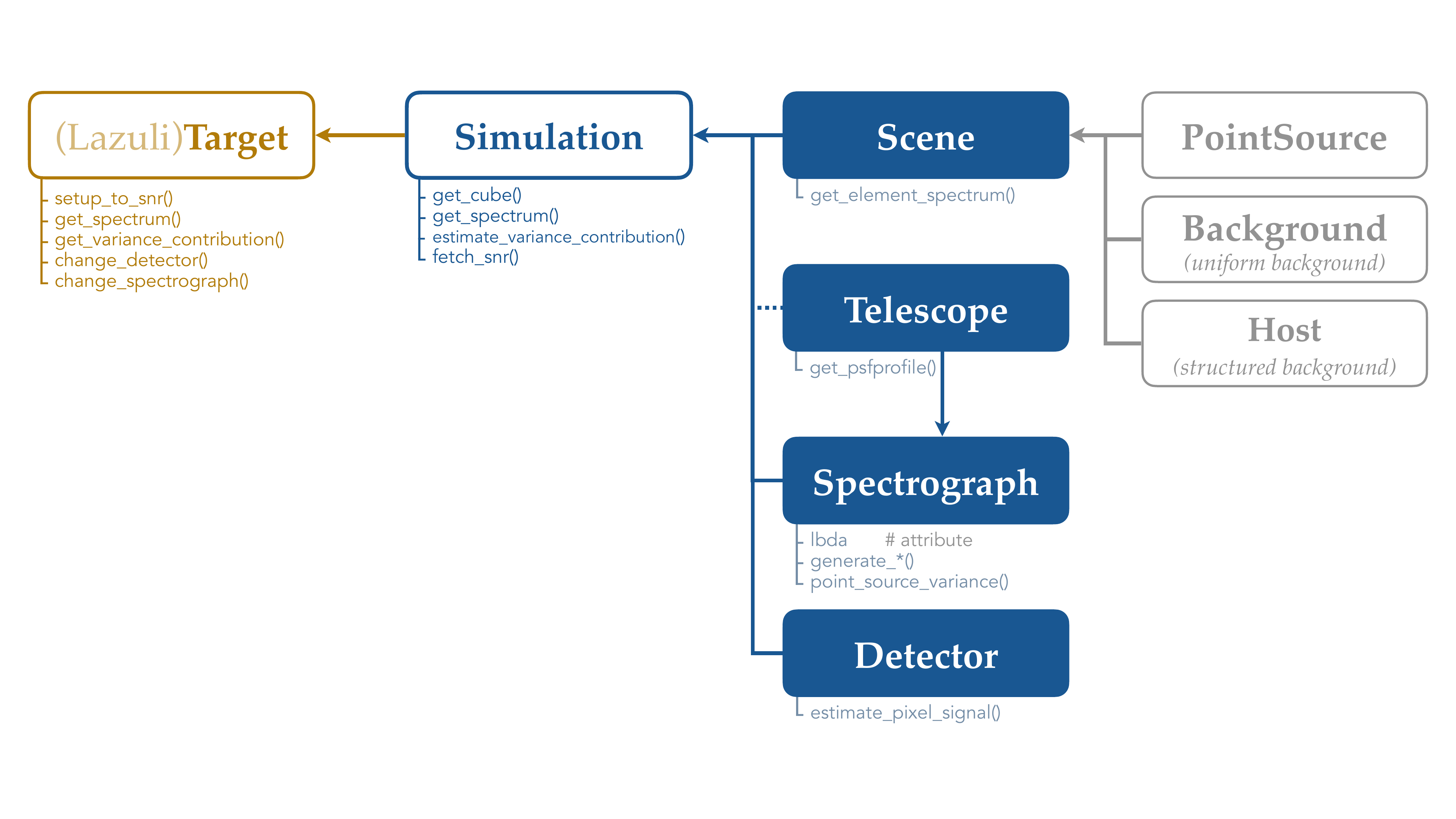}
  \caption{Illustration of the \texttt{slicersim} code structure. Elements in boxes are classes, and arrows show the connections between them, such that a \texttt{Scene} has a \texttt{PointSource}, a \texttt{Background} and a 
 \texttt{Host} instance as attributes. The \texttt{Simulation} is the cornerstone class of the package and \texttt{Target} is a top level class to simplify the user experience. The \texttt{LazuliTarget}, which inherits \texttt{Target} provides additional Lazuli specific functionalities. Some key methods or attributes are listed below these objects for illustration.}
  \label{fig:code_structure}
\end{figure*}

To simulate realistic IFS observations, one has to model the complex interplay between the astrophysical scene of interest, the telescope, the spectrograph, and the detector. Consequently, the \texttt{slicersim} package is built around these four components: the \texttt{Scene}, the \texttt{Telescope}, the \texttt{Spectrograph}, and the \texttt{Detector}. These components are brought together in the \texttt{Simulation} class, which orchestrates the simulation of an observation. We will review each component in the following sub-sections. The code infrastructure is illustrated in Fig.~\ref{fig:code_structure}.

For convenience, a suite of \texttt{Target} classes has been implemented to simplify user interaction with \texttt{Simulation}s through high-level methods, such as building a specific scene (i.e., a Type~Ia Supernova or a CALSPEC standard star), accessing the exposure time calculator (ETC), probing and visualizing the variance contributions. 
Functionalities or features specific to an instrument are developed as dedicated target subclasses. For instance, the current version of \texttt{slicersim} is released with \texttt{LazuliTarget}s, which inherit from the generic \texttt{Target}, but has, in addition, tools specific to Lazuli, such as the management of the  two spatial channels of this IFS, or high-level functionalities related to the IFS detector (see Section~\ref{sec:comp_spectro_projecttodetector}).

\subsection{\texttt{Scene}}
\label{sec:comp_scene}

\texttt{Scene} is the class that simulates the astrophysical observation as given by nature. It is itself a container that relies on three lower-level elements (python classes): the \texttt{PointSource}, a spatial point source with sky location and a spectrum; the \texttt{Background}, a spatially uniform spectral source; and the \texttt{Host}, a spatially structured background with (potentially) spatially varying spectrum.

In the way \texttt{slicersim} is built, each of these elements uses methods to define their spectra. These functions take a \texttt{wavelength} as required argument and may have additional key-arguments. For example, a SN~Ia \texttt{PointSource} model can be generated using the Spectral Adaptive Lightcurve Template, aka SALT, model \citep{guy2007, guy2010,betoule2014, kenworthy2021}, and hence the function accepts in addition to wavelength, a phase, a redshift, a stretch, a color, and an amplitude.

In version 1.0, \texttt{slicersim} has the following built-in elements:

\begin{itemize}
 \item \slobject{PointSource}: For any input spectrum (wavelength and flux), \texttt{slicersim} will build an interpolator to enable a simulation of the IFS spectrum as observed. It could also be a SN~Ia (either SALT or twin-embedding, \citealt{boone2021, ganot2025}); any \href{https://sncosmo.readthedocs.io/en/stable/source-list.html}{\texttt{sncosmo}} source \citep{barbary2023}, which includes a large selection of core collapse supernovae templates; a black-body spectrum; any CALSPEC source. 
 \item \slobject{Background}: \texttt{slicersim} considers the zodiacal light, using the \cite{aldering2001} model as implemented by Euclid \citep[see section 5.1.1 of][]{scaramella2022}.
 \item \slobject{Host}: Currently, the sole host galaxy implementation under development in \texttt{slicersim} is a simple Sersic profile with spatially uniform spectrum. Future implementation will contain more realistic models and shall accept any 3D galaxy cube (RA, Dec, $\lambda$), e.g., as given by \texttt{GalSim} \citep{rowe2015}.
\end{itemize}
\mr{As documented in the corresponding functions and methods, fluxes are expected to be given in $\mathrm{erg/s/cm^2/A}$ and wavelengths in Angstrom ($\AA$). More functionalities are in development, such as based on zero-points, see online documentation for up to date information.}

The \slobject{Scene} class has these three elements as attributes and has methods to generate the spectro-spatial scene as given by nature, i.e., prior to being affected by the point-spread function (driven by the telescope optics in the case of a space telescope). A \texttt{Scene} instance needs an input wavelength array to generate the scene spectra. This is provided through the \slobject{Simulation} object, which grabs it from its \texttt{spectrograph} attribute (see Section~\ref{sec:comp_simulation}).

\subsection{\texttt{Telescope}}
\label{sec:comp_telescope}

The \slobject{Telescope} is a simple optical object composed of mirrors. It is defined by the entrance pupil diameter and, any central obscuration if present, along with the temperature and thermal emissivity of all mirrors.

The geometry of the primary mirror (and central obscuration) defines the diffraction pattern that will affect the \slobject{Scene} prior to it reaching the spectrograph. In its released version \texttt{v1.0.0}, \texttt{slicersim} does not integrate a module for a PSF perturbed by the atmospheric as its primary usage is for the Lazuli Space Observatory.

Lazuli is designed to have an unobscured optical train with diffraction limited image quality at $500$~nm \citep{roy2026}, and the associated PSF can hence be modeled by an Airy disk. 
To account for the telescope jitter, this spatial 3D-PSF is convolved by a 2D Gaussian distribution; this scatter can be set to 0, but the target  
the Lazuli jitter requirement of $<10$~mas (root mean square), which will only slightly enlarge the native PSF. 

The optical elements of Lazuli radiate a blackbody spectrum that will completely overlap with the target light reaching the spectrograph. This signal is the sum of the thermal emission produced by the Lazuli \slobject{Telescope} and spectrograph optical elements \mr{(prior to the disperser)}, each having its own temperature and emissivity; see details on the thermal model in section~\ref{sec:model_thermal}.  Since it directly builds on top of the \slobject{Scene}, it can be treated as an additional  contribution to \slobject{scene.background}. Unlike the thermal emissions from optical elements of the \texttt{Spectrograph} following the disperser element and close to the detector, these thermal self emission (TSE) signals get dispersed within the spectrograph as for astrophysical targets.

\subsection{\texttt{Spectrograph}}
\label{sec:comp_spectrograph}

The \slobject{Spectrograph} (the IFS instrument) is a central component of the simulation. It defines how the sky is split into sub-spatial elements and how the wavelengths are dispersed. The ``slicer'' spectrograph design is fully supported by \texttt{slicersim} (\texttt{SlicerSpectrograph}). In addition, a micro-lens array (MLA) design has also been developed (\texttt{MLASpectrograph}), mostly for testing and comparison purposes, though not all functionalities are compatible with this extra design (e.g., the projection onto the detector, see Section~\ref{sec:comp_spectro_projecttodetector}). 
Both designs inherit from the same \texttt{Spectrograph} class and vary only by how the light is spatially split and the need of a cross-dispersion profile for MLA spectrographs. The Lazuli IFS spectrograph uses a slicer.

A \slobject{Spectrograph} needs three inputs: a (chromatic) spectral resolution (e.g. the resolving power, $R$, or equivalent), a (chromatic) throughput, and ``spaxel'' configurations. It also accepts settings for the thermal properties of optical elements, which are assumed to be null if not otherwise specified.

As illustrated in Fig.~\ref{fig:code_structure}, a \slobject{Telescope} instance is (optionally) passed to the \slobject{Spectrograph} to simplify the code, as it holds the information about the spatial PSF entering the spectrograph (including the jitter), which is used to generate noiseless cubes; see Section~\ref{sec:comp_spectro_getcube}.

\subsubsection{Spectral resolution}
\label{sec:spectralresolution}

The resolving power $R$ is defined by the ratio between the wavelength $\lambda$ and spectral resolution element $\Delta \lambda$, such that $R(\lambda)\equiv\frac{\lambda}{\Delta \lambda}$. From this quantity, \texttt{slicersim} can build a dispersion curve, i.e., the (wavelength dependent) step(s) on the detector between two consecutive wavelength bins ($\delta \lambda$). This assumes a dispersion resolution or \textit{wavelength solution}, which corresponds to the number of detector pixels per $\Delta \lambda$ values as a function of wavelength. The dispersion resolution is a mutable spectrograph parameter, set to 2 by default (Nyquist limit); \mr{see Table~\ref{tab:lazuliparams} for parameters used in the context of Lazuli.}

In practice, \texttt{slicersim} also directly accepts a dispersion curve (array), from which it reconstructs the resolving power, $R$, assuming a wavelength range of definition and a dispersion resolution.

\subsubsection{Spaxels}
\label{sec:spaxels}

A ``spaxel'' is the spatial element on the sky as seen by the spectrograph \citep{cappellari2003}. 
The decomposition into spaxels happens in two steps for a slicer. 
First, $n$ mirrors split the sky in one direction, defining $n$ rectangular ``1D'' sub-images. Its narrow dimension defines the width of the slicer spaxel. The long one of the slice is sampled by the detector.
To ensure equal sampling of the slit and the spatial direction at the detector, it is the norm for a slicer to have an anamorphic magnification of 2:1, such that the PSF is sampled the same by the slicer mirrors as it is by the detector pixels along the spatial direction. \mr{Indeed, the narrow dimension (the width of the slice) contains 2 detector pixels to sample the line spread function.}

\subsubsection{Throughput}
\label{sec:throughput}

The throughput curve, $\mathcal{T}$, defines the fraction of photons that reach the detector after going through the telescope and the instrument. The detector quantum efficiency (\attribute{qe}, see Section~\ref{sec:comp_detector}) is not included as part of the throughput in \texttt{slicersim} since the HgCdTe type detector baselined for Lazuli will be non-linear and have a quantum yield that can be greater than unity. 

For convenience, \texttt{slicersim} has a variety of implementations to allow the user to choose how to set throughput. 
It could be either a simple float (wavelength independent), a ``spectrum'' (wavelength and throughput) or a throughput per optical component, each having its own number of elements (\attribute{noptics}). If the latter is given, the effective ``spectrum'' is computed on the fly as $\mathcal{T}(\lambda)=\prod_i \mathcal{T}_i^{\mathrm{noptics}_i}(\lambda)$. 
This more complex implementation enables testing the impact of the coating or the number of mirrors on the overall performance of the spectrograph. The object \slobject{OpticalThroughput}, generated by \slobject{Spectrograph}, handles this functionality. 
The telescope throughput is incorporated as part of the spectrograph throughput by default, but alternative implementation are accepted by the package.

\mr{In the Lazuli configuration provided with this package, the throughput is given as the effective total throughput (i.e. as a "spectrum") for each of the two Lazuli fields; see Fig.~\ref{fig:lazuliprop}.}

\subsection{\texttt{Detector}}
\label{sec:comp_detector}

The \slobject{Detector} sets how the data is recorded and consequently how the signal variance is built. The \slobject{Spectrograph} models how the incident irradiation (scene + telescope thermal radiation) is dispersed onto the detector and \slobject{Detector} computes how each pixel reacts and records this incoming flux. 
The Lazuli spectrograph will use a Teledyne H4RG-10 detector, which generates signal ramps from non-destructive read-outs. In the released version of \texttt{slicersim}, only HxRG-like CMOS detectors have been implemented that feature non-destructive, sample-up-the-ramp (SUTR) read modes; CCD-like detectors are on the list of future implementations.

An (HxRG) \slobject{Detector} is defined by: its pixel size (\attribute{pixel\_size}), the array shape (\attribute{shape}), the time between two frame reads (\attribute{tframe}), a dark current (\attribute{dark}), a per frame read-out noise (\attribute{ron}), a (wavelength dependent) quantum efficiency (\attribute{qe}), a gain (\attribute{gain}), a saturation level (\attribute{saturation}) and a read-out mode (aka MACC mode) \attribute{nmd} specifying ``n'' groups, ``m'' frames per group, and ``d'' drops between groups. We also include the term \attribute{roic\_glow} to account for glow from the readout integrated circuit (ROIC) due to the in-pixel source-follower \citep{ReganBergeron2020, Ives2020}, which acts like a per-frame-read dark current ($e^-/\rm{frame}$; see parameter units in Table~\ref{tab:lazuliparams}).

A ``ramp'' is defined between two detector resets. 
In \texttt{slicersim} one can set a maximum number of groups allowed per ramp, such that the multi-ramps mode is used for faint targets that would need more acquisition time than that allowed by this maximum number of groups. Alternatively, this mode could be used to acquire time series of integrations.

Given the use of \attribute{nmd}-ramp to decrease the effective read noise, the variance of each pixel could be computed either using the \cite{raucher2007} model (as updated by \cite{raucher2010}, default) or by that from \cite{kubik2016} (as updated by \citealt{cogato2026}). Both agree within a few percent.

More complex detector effects that affect variance, such as pixel correlations (e.g. inter-pixel capacitance) are ignored in the current version. \mr{A detailed detector simulator (\texttt{pyxim}, Lacroix et al. in prep) is under development and eventually will be interfaced with \texttt{slicersim}.}


Altogether, given an incident $\mathrm{flux(\lambda)}$ (in photon/s/pixel), the code estimates a signal and a variance. 
The 3D variance cube is built (in ADU$^2$) following the aforementioned models, which input a combination of the flux, the read-out noise, the dark (including the ROIC glow), the frame duration and the read-out mode; see details in \cite{kubik2016}, \cite{cogato2026}, \cite{raucher2007} and \cite{raucher2010}. The 3D signal cube (in ADU) is simply built following this equation (for each spatial element $ij$):
\begin{equation}
\label{eq:flux_detector}
\begin{split}
     \mathrm{flux}_{ij}(\lambda)_\mathrm{[adu]} = &\left( \mathrm{flux}_{ij}(\lambda)_{[ph/s]} \times \mathrm{\attribute{qe}}(\lambda)_{[e^-/ph]} + \mathrm{background}_\mathrm{tot\, [e^-/s]} \right) \times\\
     & t_\mathrm{int\, [s]} \times \mathrm{\attribute{gain}}_{[adu/e^{-1}]},
 \end{split}
\end{equation}
$\mathrm{background}_\mathrm{tot}$ representing the sum of the dark current from the detector (\attribute{dark}) and the (undispersed) thermal emission from optical elements within the spectrograph (\attribute{thermal\_dark}; see Section~\ref{sec:model_thermal}); $t_\mathrm{int}$ is the effective integration time defined as $(n-1)\times(m+d)\times\mathrm{\attribute{tframe}}$ for a given the \attribute{nmd} read-out mode.

\subsection{Thermal model}
\label{sec:model_thermal}

The thermal signal model is induced by thermal radiation from the hardware elements. 
For each, the thermal signal spectrum is defined as:
\begin{equation}
    f_\mathrm{thermal}(\lambda, T) = \epsilon \times f_\mathrm{bb}(\lambda, T) \times \Omega \times s  \times n 
\end{equation}
where $\epsilon$ the element's emissivity and $n$ the number of optics per element; 
$s$ is the effective area (in $m^2$) and $\Omega$ the solid angle in steradians \mr{(solid angle of the spaxel for optics contributions prior to the disperser or of the pixel for post-disperser contributions; see below.)}.
The thermal signal itself is assumed to be a blackbody spectrum at the optic's temperature (in $K$). In \texttt{slicersim}'s thermal model functions, if the detector $\attribute{qe}(\lambda)$ is given, the thermal flux could be converted into an electron flux [$e^{-}/\mathrm{s}/\mathrm{sr}/\mathrm{m}^2/\AA$] rather than a (default) photon flux [$\mathrm{ph}/\mathrm{s}/\mathrm{sr}/\mathrm{m}^2/\AA$]. 
When a structure is made of several elements (e.g., the telescope has several mirrors), the effective signal is the sum of each individual contribution. In such cases, we assume a constant (unique) solid angle.

For the Lazuli IFS,  we have two sources of thermal emission: \textit{i}) the telescope and the spectrograph optics before the disperser, and \textit{ii}) the IFS optics between the disperser (including the prism itself) and the detector. The former is dispersed and thus effectively is similar to a ``background'' scene signal, while the latter is undispersed, and thus acts more like a detector dark. Consequently, to compute the impact of this undispersed thermal emission, we include the wavelength-dependent \slobject{Detector}.\attribute{qe} to get this effective pixel dark in electrons.
For the telescope mirror and instrumental optics prior the IFU, $\Omega$ is given by the spaxel sky area and $s$ is the primary mirror surface area.  For optics within the spectrograph, the solid angle is given by the optics' f-ratio (and shape, e.g., circular or rectangle) and $s$ by the physical detector pixel size. 

\subsection{\slobject{Simulation}}
\label{sec:comp_simulation}

\slobject{Simulation} gathers \slobject{Scene}, \slobject{Telescope}, \slobject{Spectrograph} and \slobject{Detector} objects together as attributes and has methods to interact with these, change their parameters while insuring self-consistency and pass information as needed to obtain a realistic IFS simulated observation.

In particular, the \slobject{Simulation} object passes the wavelength array (\attribute{lbda}) generated by \slobject{Spectrograph} to \slobject{Scene} to create the noiseless scene cube and to 
\slobject{Detector} to get its wavelength dependent properties, such as the \attribute{qe}. One can also obtain the effective total transmission, which combines the \slobject{Spectrograph} throughput curve with the detector \attribute{qe} and \attribute{gain}. 

In the following subsections, we detail how \slobject{Simulation} is used to generate a realistic (noisy) cube (\ref{sec:comp_spectro_getcube}), 
to estimate a realistic spectrum (\ref{sec:comp_spectro_getspec})
or to predict the expected projected spectrograph detector image
(\ref{sec:comp_spectro_projecttodetector}).

\subsubsection{Generating a realistic cube}
\label{sec:comp_spectro_getcube}

The core purpose of \slobject{Simulation} is to generate a realistic \attribute{cube} (\method{get\_cube()}), which returns both the cube ``as observed'' and the variance cube. 

This cube generation works in two steps. 
First, a noiseless cube is generated (\method{get\_projected\_scene()}). To do so, the \slobject{Scene} spectra (1D array for \slobject{PointSource} and \slobject{Background}, a 3D-array for \slobject{Host}) is generated given the \slobject{Spectrograph}'s wavelength and is then convolved by the effective ``spatial$+$jitter'' PSF from the \slobject{Telescope}; \mr{in current the implementation of \texttt{slicersim} the diffraction induced by the slit width itself is ignored.}
Once the cube is generated, the effective \slobject{Spectrograph}'s LSF is applied along the wavelength direction. In \texttt{slicersim v1.0.0} all spectra have the same LSF, which effectively means that the spectrograph-induced PSF is uniform across the detector. A more realistic behavior is currently under development and will be included in a near future release.

Second, we generate the variance cube associated with this noiseless cube. This step differs for ``MLA'' and ``slicer'' IFU designs within \texttt{slicersim}. 
In a slicer, the data cube directly projects onto the detector. This means that each cube element ($\mathrm{flux}(i,j,\lambda)$) is projected onto the detector and is thus directly affected by the detector noise as described in section~\ref{sec:comp_detector}.

For an MLA-based IFU, one has to first compute the wavelength-dependent cross-dispersion profile to estimate how many detector pixel a $\mathrm{flux}(i,j,\lambda)$ cube element is effectively projected onto. The current release assumes a Gaussian cross-dispersion profile, whose width is linearly increasing with wavelength. The cube ``as observed'' is drawn from the noiseless cube given the variance cube assuming Gaussian noise.

\begin{figure}
  \sidecaption
  \includegraphics[width=\linewidth]{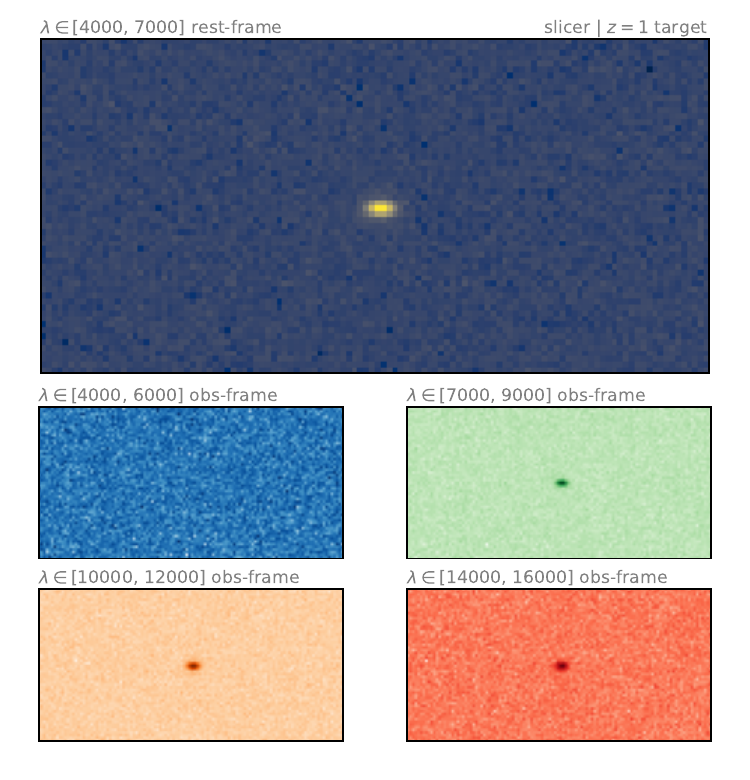}
  \caption{Illustration of a simulated slicer cube observing a host-less $z=1$ Type~Ia Supernovae observed using two \attribute{nmd=(64, 8, 0)} H4RG ramps (48 min). This assumes the Lazuli IFS narrow-field (2.3" x 4.6" field of view) current best estimate configuration (see Table~\ref{tab:lazuliparams}). 
  \emph{Top}: Integrated cube along the wavelength axis between 400 and 700~nm rest-frame. 
  \emph{Bottom}: Integrated cube in four difference observer-frame wavelength ranges (see titles). 
  }
  \label{fig:sceneslice}
\end{figure}

In Fig.~\ref{fig:sceneslice}, we illustrate the cube generation by a slicer spectrograph observing a $z=1$ SN~Ia.
\mr{In this example, we notice that the PSF appear elongated along the slice length (x-axis direction). This  is caused by instrumental aberrations (see sampling Table~\ref{tab:lazuliparams}) that blur the signal past the IFU (slicer plane). Hence, since the spatial sampling along the slice length is made at the pixel level, this aberration mimic a spatial PSF (and an line-spread function, not visible here); see Fig.~\ref{fig:mapping} for an illustration that ignores instrumental aberration. In this illustration, the aberrations add a typical 1 pixel extra scatter, which is what would be expected for Lazuli.}

The bluest part of the cube has no significant signal, as the target is highly redshifted and SNe~Ia emit little flux at rest-frame UV wavelengths (see extracted spectrum in Fig.~\ref{fig:spectra}). When integrated over large wavelength ranges (200~nm bins in this example) the point source is significantly detected, but the signal-to-noise per spectral resolution element is ``only'' $\sim8$ ($\sim5$ per wavelength bin) for this example (see Fig.~\ref{fig:variance_contribution}). We finally notice in this figure that the point spread function gets larger with wavelength, as expected for a diffraction limited instrument.

\subsubsection{Extracting the target spectrum}
\label{sec:comp_spectro_getspec}

The target spectrum and its variance is usually the final product of interest to a the user. These are accessible using the \method{Simulation.get\_spectrum()} method.

Once the realistic cube and its associated variance cube have been generated (see Section~\ref{sec:comp_spectro_getcube}), the expected variance spectrum is extracted. In \texttt{slicersim 1.0} this extraction is optimal: we use the same PSF to generate the cubes and to estimate the variance spectrum (from the variance cube). 
We then take the spectrum from the \slobject{Scene} (so not from the realistic cube), which is then randomly sampled given the variance spectrum assuming Gaussian noise. 
To study biases, one would need to extract the spectrum for the realistic cube after perturbing it by the biases of interest. This is a planned development beyond the released version 1.0.

\begin{figure}
  \sidecaption
  \includegraphics[width=\linewidth]{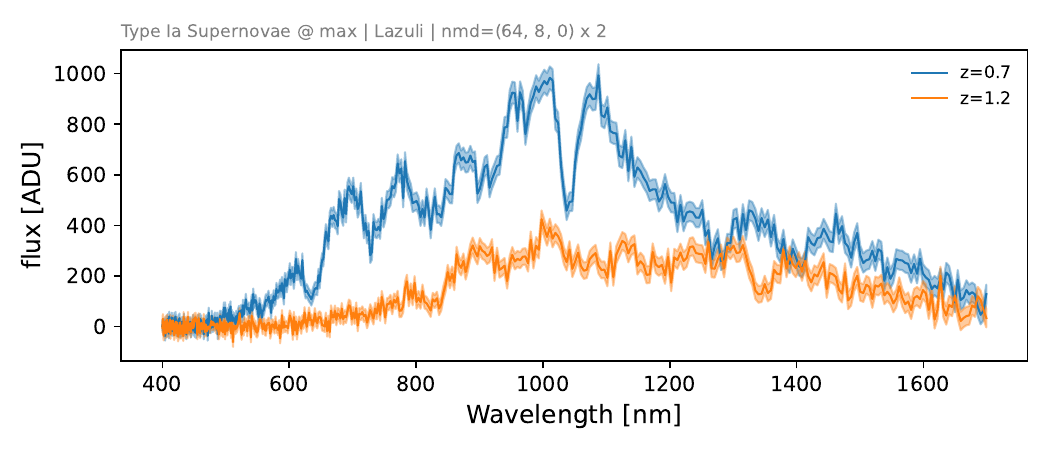}
  \caption{Two Type~Ia Supernovae spectra ($z=0.7$, in blue; $z=1.2$, in orange) simulated as observed with the Lazuli narrow-field slicer using two \attribute{nmd=(64, 8, 0)} H4RG ramp, corresponding to a 2900~s exposure time. For the $z=1.2$ case see an illustration of the cube in Fig.~\ref{fig:sceneslice} and variance details in Fig.~\ref{fig:variance_contribution}.}
  \label{fig:spectra}
\end{figure}

We present in Fig.~\ref{fig:spectra} two Type~Ia Supernovae spectra simulated with the SALT model at maximum light assuming a stretch and color of 0, i.e. a typical target. These spectra where simulated assuming the Lazuli narrow field slicer and setting the detector read-out mode to two \attribute{nmd=(64, 8, 0)} ramps, which corresponds to an exposure time of 2900~s. 
With this configuration, we obtain an average signal-to-noise ratio per wavelength bin in the $[600, 700]$nm rest-frame wavelength range of 16 for the $z=0.7$ target and of 5 for the $z=1.2$ case. For convenience, we implemented the \method{Simulation.get\_band\_snr()} method to easily provide such information.

\subsubsection{Generating the detector image}
\label{sec:comp_spectro_projecttodetector}

One does not need to project data onto the detector to precisely estimate the expected signal variance and, thereby, build a realistic signal-to-noise estimator. However, having the possibility to do so is convenient for visualizing how the data may actually look and for building the inference pipeline that will start from real data to measure the input flux.

To build the expected detector image, \texttt{slicersim} requires external input --- often called an ``optical model'' --- to build the mapping function. This function converts $(x, y, \lambda)$ coordinates at the slicer (or MLA) plane into $(i, j)$ coordinates at the detector plane;  $x$, $y$ being the sky-coordinates relative to the center of the field of view.
In practice, \texttt{slicersim} only needs a few samples per slice (lenslet element) and performs a 2D interpolation. Once this model is built, the package can also provide the inverse function $(i, j)$ to $(x, y, \lambda)$. This mapping is handled by the \slobject{SlicerMapper} objects.

\begin{figure*}
  \sidecaption
  \includegraphics[width=\linewidth]{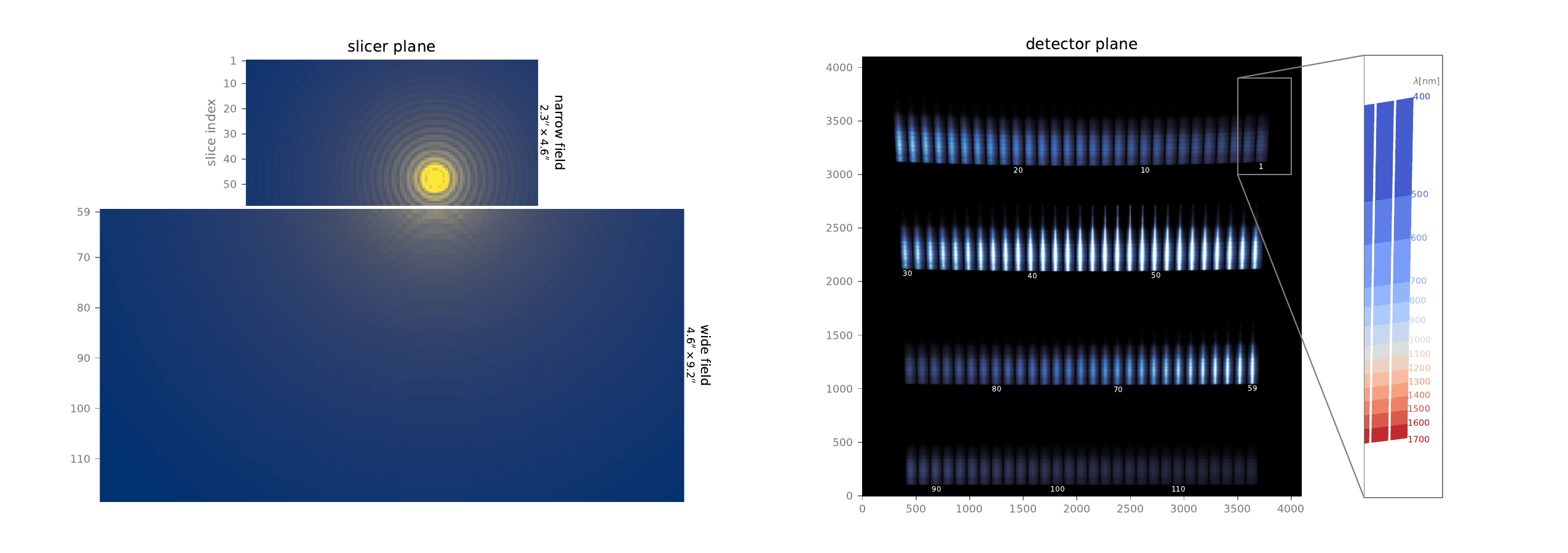}
  \caption{
  \mr{Illustration of the Lazuli two-channel slicer data. In this illustration aberrations from the spectrograph have been removed to illustrate the incoming PSF (airy) and background contributions (host galaxy, astrophysical or thermal) have been switched off to avoid signal offset between the two channels.
  The \emph{left} panel show the 2D scene averaged between 1600~nm and 1700~nm, as projected onto the slicer plane. The top panel corresponds to the narrow field (where the core of the target is located), while the bottom panel is the wide field with twice large slices (40mas vs. 80mas, see Table~\ref{tab:lazuliparams}).  
  Each slice is then mapped onto the detector as indicated by the slice number (1 to 116), matching between the slicer plane (\emph{left}) and the detector (\emph{right}). We see that the narrow field projects into the upper half of the detector and that central slices ($\sim30$) are located on the left while outer slices ($\sim0$ or $\sim58$) are on the right. This same is true for the wide field, but on the lower-half  of the detector.
  A zoom on the first slices is shown in the \emph{right-most} inset where we replaced data by wavelength projection to illustrate how light is dispersed by the spectrograph. Each color panel corresponds to 100~nm. Notice that the bluest wavelength are projected upward and span a large detector area, as they have a higher dispersion (see Fig.~\ref{fig:lazuliprop}).}
  }
  \label{fig:mapping}
\end{figure*}

This mapping is illustrated in Fig.~\ref{fig:mapping} for Lazuli. This \mr{observatory} IFS is designed to have two observing fields: a narrow and a wide field. 
To illustrate the point-source scenario, this figure shows a pure ``point-source'' cube \mr{(no host galaxy, no background) and the instrumental aberrations have been turned off to clearly see the incoming PSF}. 
Since Lazuli is a slicer spectrograph, slices are located side by side on the detector. \mr{The narrow (wide) field projects slices on the top (bottom) half of the detector, with slices located at the center of the field of view, near \#30 (\#90) are located near the left edge of the detector. For each slice, the dispersion goes downward with increasing wavelengths. This figure further illustrates the resolving power of the spectrograph, that is much higher in bluer wavelength (see also Fig.~\ref{fig:lazuliprop}. In this figure this is visible as blue wavelengths span a much large number of pixels for a fixed wavelength binning than red ones.}

Such external data may be derived from a Zemax optical model (as for Fig.~\ref{fig:mapping}). The optical model itself will likely be refined with on-sky calibration.

\subsection{\slobject{Target} and \slobject{LazuliTarget}}
\label{sec:comp_lazulitarget}

The \slobject{Simulation} object is a powerful tool that allows one to interact with all aspects of the code. As such, it could be quite complex to use for non-expert users. Hence, the \texttt{Target} object has been designed to simplify the user experience, focusing on top level functionalities and simplifying the initialisation of common scenes like that of a star or of a Type~Ia Supernova.

A \texttt{Target} has a \slobject{Simulation} instance attribute upon which all  top-level methods are implemented, such as changes to the spectrograph or detector parameters (\method{target.change\_...()}), accessing data (\method{target.get\_cube()} or \method{target.get\_spectrum()}), accessing the exposure time (\method{target.get\_exptime()}) or to setting the detector properties to reach a requested signal to noise ratio \method{target.setup\_to\_snr()}; see usage details in Section~\ref{sec:model}.

\texttt{slicersim} contains several pre-defined \texttt{Target}s: \slobject{Supernova} to load and change supernovae (any kind) scenes; \slobject{CalSpec} to access any CALSPEC star simply by its name, while the generic \slobject{Target} accepts any spectrum as input. All inherit from the same \slobject{VirtualTarget} virtual class that holds most of the methods and attributes.

To provide yet another level of simplicity, we have developed the \texttt{LazuliTarget} object, which inherits from the generic \texttt{Target} but with additional functionalities specific to the Lazuli IFS \citep{roy2026}. For instance, Lazuli's IFS contains two slicer fields, \texttt{LazuliTarget} hence contains methods associated with this specificity, e.g., by overloading the \method{lazulitarget.change\_spectrograph()} method. There is a \texttt{LazuliTarget} for each of the aforementioned predefined targets.

\begin{figure}
  \centering
  \includegraphics[width=\linewidth]{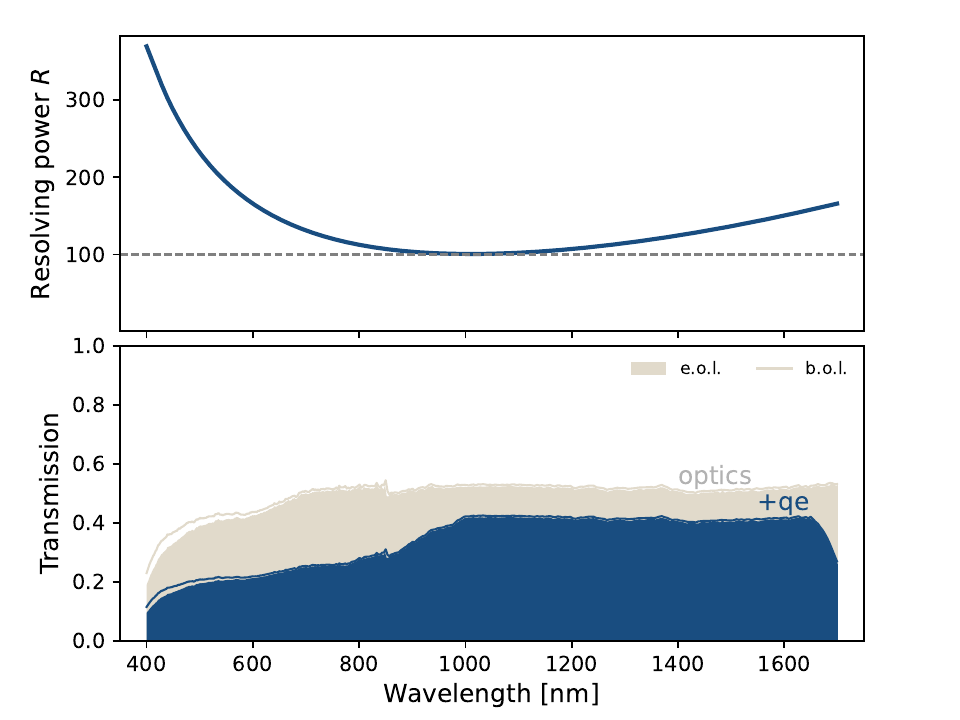}
  \caption{Illustration of default assumptions for the Lazuli IFS simulations; end-of-life (e.o.l) and beginning-of-life (b.o.l). \emph{Top}: Spectrograph resolving power ($R=\frac{\lambda}{\Delta \lambda}$). \emph{Bottom}: Absolute  transmission of the telescope plus IFS instrument, then that combined with the detector quantum efficiency to give the net IFS efficiency. Performances are expected to only slight decrease from beginning to end of life operations, and mostly in bluest wavelengths. See configurations in table~\ref{tab:lazuliparams}.
  }
  \label{fig:lazuliprop}
\end{figure}

These Lazuli specific objects automatically load either \mr{the beginning-of-life / current best-estimate (default) or the end-of-life / conservative  configuration parameters} 
listed in Table~\ref{tab:lazuliparams} and illustrated in Fig.~\ref{fig:lazuliprop}. 
As such, loading a realistic simulation is as simple as \texttt{snia = slicersim.LazuliSupernova(redshift=1)}.
The assumed default spectrograph resolving power, effective optical throughput, and detector quantum efficiency are illustrated in Fig.~\ref{fig:lazuliprop} \mr{for both eol and bol. As illustrated in this figure, performance of the instrument should only slightly vary and mostly in the bluest wavelengths.}
We highlight nonetheless that while these curves are realistic and close to the expected instrumental performances, they are not the actual values of the as-built instruments, which would be updated for future \texttt{slicersim} releases as those numbers become available.

\begin{table}
\centering
\tiny
\caption{Main Lazuli IFS simulation default parameters.}
\label{tab:lazuliparams}
\begin{tabular}{l c c c}
\hline\\[-0.8em]
parameter name  & b.o.l$^\dagger$ & e.o.l$^\dagger$ & units\\[0.30em]
\hline\\[-0.8em]
\hline\\[-0.5em]
Detector & &\\[0.30em]
\hline\\[-0.5em]
dark                    &   2.8e-3      &   1e-2                                    & e-/s/pix      \\[0.15em]
roic glow               &   3e-3        &   1e-2                                    & e-/s/pix      \\[0.15em]
read-out noise     &   10          &     20                                    &    e-/frame      \\[0.15em]
gain                    &   \multicolumn{2}{c}{1                 }                  &    ADU/e-        \\[0.15em]
saturation              &   \multicolumn{2}{c}{65 535             }                 &    ADU           \\[0.15em]
frame time              &   \multicolumn{2}{c}{2.83              }                  &    s             \\[0.15em]
qe                      &   \multicolumn{2}{c}{``\emph{H4RG17}'' }                  &    e-/ph  \\[0.15em]
\hline\\[-0.5em]
Spectrograph & & \\[0.30em]
\hline\\[-0.5em]
slice width             &   \multicolumn{2}{c}{(40 or 80)$^\ddagger$        }                  &    mas           \\[0.15em]
slicer layout           &   \multicolumn{2}{c}{$58 + 58^\ddagger$      }                  &    --            \\[0.15em]
slicer FoV           &   \multicolumn{2}{c}{$(2.3"\times 4.6") + (4.6" \times 8.8")^\ddagger$      }                  &    --            \\[0.15em]
anamorphism             &   \multicolumn{2}{c}{$2\times1$        }                  &    --            \\[0.15em]
dispersion curve        &   \multicolumn{2}{c}{``\emph{dispersion\_offner}''}       &   pixel \\[0.15em]
dispersion resolution   &   2.3          &     2.5                                  &    pixel         \\[0.15em]
throughput              &   ``\emph{throughput\_bol}'' & ``\emph{throughput\_eol}'' &   --        \\[0.15em]
optics: emissivity      &   \multicolumn{2}{c}{ [0.05, 0.05, 0.03, 0.03, 0.03]}     & --     \\[0.15em]
optics: n-elements      &   \multicolumn{2}{c}{ [5, (4 or 6)$^\ddagger$, 1, 1, 3]      }              &    --              \\[0.15em]
optics: temperature     &   \multicolumn{2}{c}{ [230, 228, 228, 226, 229]}               &     K          \\[0.15em]
optics: dispersed       &   \multicolumn{2}{c}{ [True, True, False, False, False]}               &     K          \\[0.15em]
\hline\\[-0.5em]
Telescope & & \\[0.30em]
\hline\\[-0.5em]
external diameter       &   \multicolumn{2}{c}{ 3                 } &    m             \\[0.15em]
internal diameter       &   \multicolumn{2}{c}{ 0                 } &    m             \\[0.15em]
guiding                 &   \multicolumn{2}{c}{ 10                } &    mas           \\[0.15em]
optics: emissivity      &   \multicolumn{2}{c}{ [0.05, 0.02, 0.02, 0.02] }& --     \\[0.15em]
optics: temperature     &   \multicolumn{2}{c}{ [293, 293, 293, 293]} &    K          \\[0.15em]
\hline
\end{tabular}
\tablefoot{$^\dagger$: b.o.l: beginning of life / best estimate; e.o.l: end of life / conservative estimate. $^\ddagger$: narrow and wide fields, respectively.
\\
Names in quotes are configuration files, giving a value as a function of wavelength. See Fig.~\ref{fig:lazuliprop}.
}
\end{table}

\section{Using \texttt{slicersim}, some more examples}
\label{sec:model}

In the following subsection, we review some key usage tools implemented in \texttt{slicersim}. 
We invite users to nonetheless check the latest package documentation for additional API features.

\subsection{Changing internal parameters}
\label{sec:update}

Each internal \texttt{slicersim} \texttt{object}, i.e., \slobject{Simulation} and below following the code structure illustrated Fig.~\ref{fig:code_structure}, has an \method{.update} method. 
This method enables the user to self-consistently change any ``mutable parameters'' of these objects. Most parameters are mutable, e.g., the read-out noise, read-out mode, dark current from the detector, the temperature and the size of the telescope mirrors, and the dispersion resolution or the spaxel size of the spectrograph; see the \attribute{.mutable\_parameter} object attribute.

Some attributes are derived parameters, such as the throughput (\attribute{.throughput}) or the wavelength (\attribute{.lbda}) for the spectrograph. They could either be changed using their root parameters (such as \texttt{noptics} for the throughput) or directly by dedicated setting methods, such as \texttt{set\_throughput}. These methods enable \texttt{slicersim} to handle more complex configuration updates than just changing a number --- the primary usage of the \method{.update} method.

The \texttt{slicersim} package has been designed such that, if the user changes a parameter, the code will remain self-consistent. 
This is why, for instance, the wavelength parameter is held solely by  the \slobject{Spectrograph} and passed to the other objects and methods through the \slobject{Simulation}. As such, any change of \attribute{spectrograph.lbda} will be naturally propagated everywhere it matters.

\subsection{Exposure times and signal-to-noise ratios}
\label{sec:fetch_snr}

\begin{figure}
  \sidecaption
  \includegraphics[width=\linewidth]{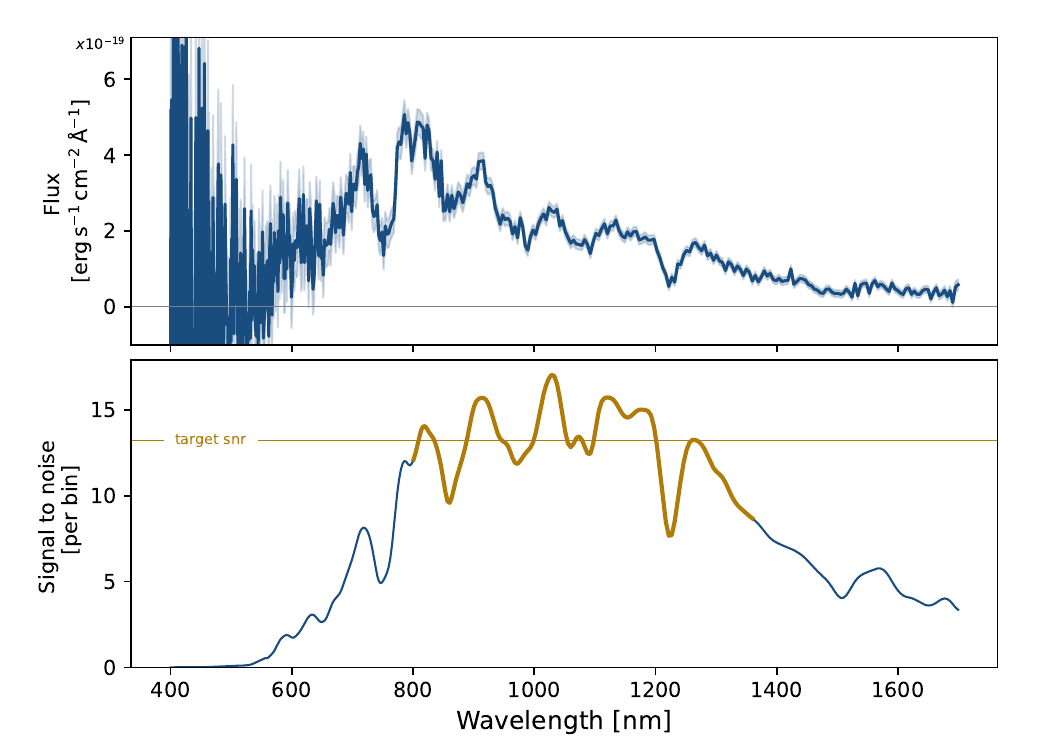}
  \caption{Illustration of a $z=1$ Type~Ia Supernova simulated in the context of the Lazuli Space Observatory integral field spectrograph. 
  \emph{Top}: Simulated spectrum requesting an average signal to noise of 20 per resolution elements in the rest-frame $\lambda\in[400, 680]$nm wavelength range. 
  \emph{Bottom}: Signal to noise spectrum. The corresponding observer-frame wavelength are highlighted in bold-orange. The horizontal line shows the signal to noise per wavelength bin corresponding to the request. In this simulation, the dispersion resolution is 2 bins, so the target SNR is $20/\sqrt{2.3}=13.2$ per bin, see text and Table~\ref{tab:lazuliparams}.
  }
  \label{fig:snr_example}
\end{figure}

The \method{.update} method (see Section~\ref{sec:update}) enables one to change the \slobject{detector.nmd} read-out mode, which effectively corresponds to changing the exposure time and consequently the signal to noise of the simulated spectrum (see Section~\ref{sec:comp_spectro_getspec}). 
One can then loop over the read-out mode configuration until one reaches the desired mean (or max, or median etc.) signal to noise in the wavelengh range of interest; be it rest- or observer-frame.
The \method{simulation.get\_band\_snr()} method simplifies the computation of the signal to noise of interest for the current configuration, while the \method{simulation.fetch\_snr()} fits for the read-out mode. These methods are illustrated in Fig.~\ref{fig:snr_example} for a $z=1$ Type~Ia supernova observed with Lazuli. Alternatively, one can directly set the read-out mode configuration (\texttt{nmd} and \texttt{nramps}) using the \method{target.change\_detector} method.

By default the \slobject{Detector} has a maximum number of 64 groups per frame and 8 frames per group; both mutable parameters. 
By default, \method{simulation.fetch\_snr()} loops over the number of groups. If, during the fitting loop the maximum number of groups allowed has been reached and the signal-to-noise is still lower than requested, the code will freeze the maximum read-out mode configuration (e.g., \attribute{nmd=(64, 8, 0)}) and then iterate over the number of ramps.
\mr{Once this has converged, if the signal to noise is too high, the number of ramps gets frozen and the number of groups per ramp is fitted; hence, in the current implementation of \texttt{slicersim}, all ramps share the same number of groups}.
Conversely, a minimum number of groups could be specified, and, upon fitting the best configuration the number of frames per group, will reduce to 4, then 1 (see option), for bright targets.

Since the individual \attribute{nmd} configurations are always identical when using the multi-ramp mode, the \method{simulation.fetch\_snr()} method simply returns both the single ramp \attribute{nmd} configuration and the number of ramps (\attribute{nramps}). In the example shown in Fig.~\ref{fig:snr_example}, we reach an average signal-to-noise of 20 per resolution elements in the $[400, 680]$nm (rest-frame) in 81 min, corresponding to four (54, 8, 0) ramps; this signal to noise converts to $\sim13$ per wavelength bins for a 2.3 pixel spectral resolution (see Table~\ref{tab:lazuliparams}).

The \slobject{Target} object uses this \slobject{Simulator} method inside its high-level \method{target.setup\_to\_snr()} method, which first fits for the optimal detector configuration (\attribute{nmd}, \attribute{nramps}) and then sets it using \method{simulation.update()}. 
Finally, one can simply use the \method{target.get\_exposure\_time()} method (or \method{simulation.get\_times()}) to obtain the total exposure time. This is how using \texttt{slicersim} as an exposure time calculator works.

\subsection{Probing the sources of variance.}
\label{sec:get_variance_origin}

Taking full advantage of the flexibility of \texttt{slicersim}, one can use the \method{simulation.update()} method to: (1) switch off any variance source, (2) compute the expected spectrum variance (\method{simulation.get\_spectrum()}), and (3) compare it with that obtained with the variance source on.
This functionality is used inside of the \slobject{simulation.get\_variance\_contribution()} method that scans  all variance sources this way.

There are currently \mr{eight} variance sources implemented in \texttt{slicersim}: the detector \attribute{dark}, read-out noise (\attribute{ron}) and ROIC glow (\attribute{roic\_glow}); the thermal emission from elements within the spectrograph after the disperser (undispersed, \attribute{thermal\_dark}); the thermal emission from the telescope and the spectrograph prior to the disperser (dispersed, \attribute{thermal}); and the photon noise from the three scene elements (\instance{pointsource}, \instance{background} and \instance{host}). 

\begin{figure}
  \includegraphics[width=\linewidth]{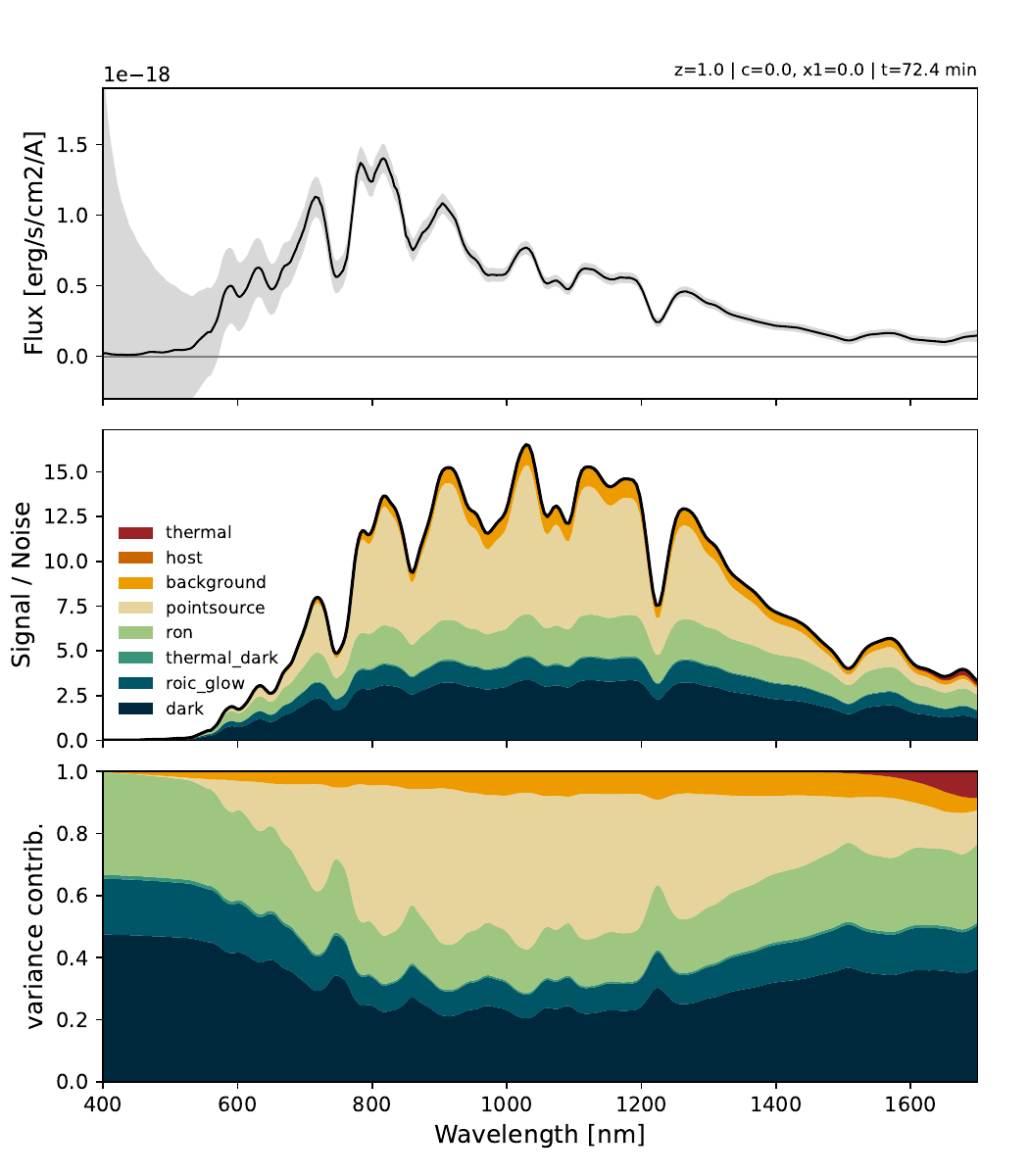}
  \caption{Details of the variance contributions for a $z=1$ Type~Ia Supernova observation acquired by the Lazuli narrow field in 72.4~min (three (64, 8, 0) ramps, as for Fig.~\ref{fig:snr_example}).
  \emph{Top}: Input target spectrum (the \instance{pointsource}) in physical units. The grey band represents the expected error for the b.o.l configuration (see Table~\ref{tab:lazuliparams}). The input spectrum hasn't been injected by the corresponding noise for illustration purposes. \emph{Middle}: signal-to-noise spectrum decomposed by variance source contribution (see legend). The full back line shows the signal to noise with all variance sources are included. \emph{Bottom}: variance contribution divided by the total variance. This figure illustrates the relative variance source impact as a function of wavelength.
  }
  \label{fig:variance_contribution}
\end{figure}

We illustrate one such variance contribution decomposition in Fig.~\ref{fig:variance_contribution} for a $z=1$ host-less SN~Ia observed with Lazuli's spectrograph with three $\attribute{nmd}=(64, 8, 0)$ ramps (as for Fig.~\ref{fig:snr_example}, Section~\ref{sec:fetch_snr}). 
As expected, the relative contribution of each source of variance varies with wavelength. 
In this example, photon noise from the target (\texttt{pointsource}) dominates the variance contribution where the flux peaks (between 8000 and 1400~nm observer-frame). 
For the Lazuli telescope, the thermal contribution from the mirrors is only visible in the reddest wavelengths but remains small in comparison to other detector effects (\attribute{dark}, \attribute{roic\_glow} and \attribute{ron}). Given that the IFS optics are all cold ($\sim230$K) the internal thermal emission (\attribute{thermal\_dark}) is also low. This nicely illustrates the use of \texttt{slicersim} to help with the conceptual design of an instrument.

\subsection{Simulating an observing campaign}
\label{sec:simulate_survey}

\begin{figure}
  \includegraphics[width=\linewidth]{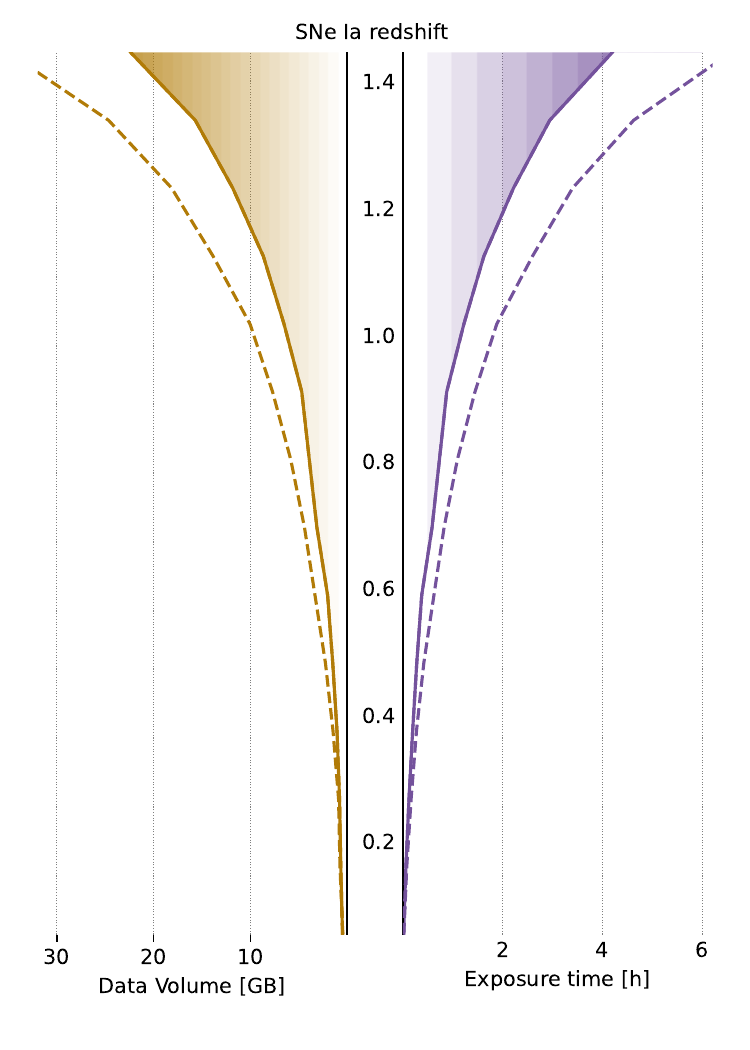}
  \caption{Illustration of the time (rightward, one shade per 0.5h) it would take to acquire a Type~Ia Supernova as a function of redshift (vertical axes) and the volume of data generated assuming an 8 frame per group H4RG grouping (leftward, one shade per GB). 
  The full lines illustrate a Lazuli simulation targeting an average signal to noise of 20 per resolution element, while dashed lines correspond to the 25 case.
  }
  \label{fig:time_and_volume}
\end{figure}

Given a list of targets and their spectral properties, one can use \texttt{slicersim} to estimate the time it would take to acquire this sample for a given signal-to-noise, and the volume of data this would generate. 
This is illustrated in Fig.~\ref{fig:time_and_volume} for a sample of Type~Ia Supernovae observed with Lazuli from $z=0$ to $z=1.5$ while requesting either an average signal to noise of \mr{20 or 25 per resolution element in the rest-frame $[400,\,680]$nm wavelength range. In this example we use the current best estimate configuration for Lazuli (b.o.l, see Table~\ref{tab:lazuliparams})} and we switch from observing with the narrow field to wide field at redshift $z_{\mathrm{switch}} = 0.8$. We do so as the SNe~Ia signal is dominating in the red to infra-red for $z>0.8$ targets (see e.g., Fig.~\ref{fig:snr_example}) \mr{and there the spatial point-spread-function starts getting over-sampled for the narrow field (see Fig.~\ref{fig:mapping}). We hence switch to the wide field to save exposure time.}

As expected, Fig.~\ref{fig:time_and_volume} shows that nearby targets are faster to observe than distant ones. If we request an average signal to noise of 20, the observing time for a $z=0.2$ SNe~Ia is 5~min, a $z=1$ takes 72.4~min (81 min using the narrow field, see Section~\ref{sec:fetch_snr}) and a distant $z=1.5$ SN~Ia requires 4.8~h. 
In comparison, targeting a signal to noise of 25, the typical exposure time increases by $\sim30\%$ up to $z=0.5$  and then closer to $+50\%$; with 6.5\,min, 97~min and 7.6\,hrs for a $z=0.2$, a $z=1$, and a $z=1.5$ SN~Ia, respectively. 
The Lazuli IFS H4RG detector would generate 23GB of grouped raw data for a $z=1.5$, while barely 1GB for a $z=0.2$ target.

With \texttt{slicersim}, we can consequently estimate the on-sky time needed for a survey to observe a given sample of targets to a desired precision. Based on the current baseline, it would take one year on-sky (i.e., 100\% of observing time) for Lazuli to observe 8000 SNe\,Ia uniformly distributed between $z=0$ and $z=1.5$, with a signal to noise of 20 per resolution elements; 1.5 years if targeting a signal to noise of 25.

We note that these these time estimates are to illustrate the capabilities of \texttt{slicersim} for survey planning, and not meant to forecast an actual Lazuli program since some instrument design parameters are not fully defined.
However, one can see how \texttt{slicersim} will bring enormous value to facilitate proper instrument \textit{and} survey design trades, resulting in the most optimal build and use of the Lazuli Space Observatory and other future missions.  

\section{Conclusion and Discussion}
\label{sec:conclusion}

We have presented \texttt{slicersim}, a new Python package for simulating IFS slicer observations. This package is a modular and flexible tool that allows the user to build realistic simulations. 
It has been developed to support the development of the Lazuli Space Observatory slicer \citep{roy2026}, but the code structure has been designed to be flexible and extensions could be made for any IFS.

The main components of the package are the \texttt{Scene}, the \texttt{Telescope}, the \texttt{Spectrograph}, and the \texttt{Detector}. These components are brought together in the \texttt{Simulation} class, which orchestrates the simulation of an observation. A top level class \texttt{Target} is made to simplify the user experience in interacting with the \texttt{Simulation}, and a set of \texttt{LazuliTarget}s are provided that automatically loads the default Lazuli properties. In this paper, we have detailed each of these components and provided an example of how to use \texttt{slicersim} as an exposure time calculator or to investigate the origin of the variance of a given observation. We have also listed nominal values for key characteristics of the expected Lazuli IFS and illustrated the use of \texttt{slicersim} in studying how such an instrument could be used to execute a Type~Ia Supernova campaign (see Perlmutter et al. in prep, \citealt{roy2026} and \citealt{wevers2026}).

\begin{acknowledgements}

\end{acknowledgements}

\end{document}